\documentclass[sigconf, nonacm]{acmart}
\usepackage{enumitem}
\usepackage{xspace}
\usepackage{tikz}
\setlist[itemize]{leftmargin=1.5em, labelsep=0.5em}
\newcommand\vldbdoi{XX.XX/XXX.XX}
\newcommand\vldbpages{XXX-XXX}
\newcommand\vldbvolume{14}
\newcommand\vldbissue{1}
\newcommand\vldbyear{2020}
\newcommand\vldbauthors{\authors}
\newcommand\vldbtitle{\shorttitle} 
\newcommand\vldbavailabilityurl{URL_TO_YOUR_ARTIFACTS}
\newcommand\vldbpagestyle{plain}

\begin{document}

\newcommand{\stitle}[1]{\noindent{\textbf{#1}}}
\newcommand{\kw}[1]{{\ensuremath {\mathsf{#1}}}\xspace}
\newcommand{\kwnospace}[1]{{\ensuremath {\mathsf{#1}}}}
\newcommand{\todo}[1]{\textcolor{red}{$\Rightarrow:$ #1}}
\newcommand{\revise}[1]{#1}
\newcommand{\delete}[1]{}
\newcommand{\notation}[1]{}

\newcommand{\update}[1]{\textcolor{red}{$\Rightarrow:$ #1}}

\newcommand*\circled[1]{\tikz[baseline=(char.base)]{
            \node[shape=circle,draw,inner sep=1pt] (char) {#1};}}

\definecolor{c1}{RGB}{42,99,172} %
\definecolor{c2}{RGB}{255,88,93}
\definecolor{c3}{RGB}{255,181,73}
\definecolor{c4}{RGB}{119,71,64} %
\definecolor{c5}{RGB}{228,123,121} %
\definecolor{c6}{RGB}{208,167,39} %
\definecolor{c7}{RGB}{0,51,153}
\definecolor{c8}{RGB}{56,140,139} 
\definecolor{c9}{RGB}{0,0,0} 

\newcommand{\sssp}{\kw{SSSP}}
\newcommand{\pr}{\kw{PR}}
\newcommand{\tc}{\kw{TC}}
\newcommand{\bc}{\kw{BC}}
\newcommand{\bfs}{\kw{BFS}}
\newcommand{\clique}{\kw{KC}}
\newcommand{\core}{\kw{CD}}
\newcommand{\cc}{\kw{WCC}}
\newcommand{\lcc}{\kw{LCC}}
\newcommand{\wcc}{\kw{WCC}}
\newcommand{\lpa}{\kw{LPA}}
\newcommand{\cd}{\kw{CD}}
\newcommand{\kc}{\kw{KC}}
\newcommand{\khop}{\kw{K}-\kw{Hop}}
\newcommand{\cluster}{\kw{Cluster}}

\newcommand{\vc}{\kw{vertex}-\kw{centric}}
\newcommand{\vcc}{\kw{Vertex}-\kw{centric}}
\newcommand{\ec}{\kw{edge}-\kw{centric}}
\newcommand{\blc}{\kw{block}-\kw{centric}}
\newcommand{\blcc}{\kw{Block}-\kw{centric}}
\newcommand{\sgc}{\kw{subgraph}-\kw{centric}}

\newcommand{\graphx}{\kw{GraphX}}
\newcommand{\power}{\kw{PowerGraph}}
\newcommand{\flash}{\kw{Flash}}
\newcommand{\grape}{\kw{Grape}}
\newcommand{\pregel}{\kw{Pregel+}}
\newcommand{\ligra}{\kw{Ligra}}
\newcommand{\gthinker}{\kw{G}-\kw{thinker}}
\newcommand{\pregell}{\kw{Pregel}}
\newcommand{\scattergather}{\kw{Scatter}-\kw{Gather}}
\newcommand{\scattercombine}{\kw{Scatter}-\kw{Combine}}
\newcommand{\xstream}{\kw{X}-\kw{Stream}}
\newcommand{\graphchi}{\kw{Graphchi}}
\newcommand{\chaos}{\kw{Chaos}}
\newcommand{\blogel}{\kw{Blogel}}
\newcommand{\arabesque}{\kw{Arabesque}}
\newcommand{\fractal}{\kw{Fractal}}
\newcommand{\autoMine}{\kw{AutoMine}}
\newcommand{\peregrine}{\kw{Peregrine}}
\newcommand{\vrdd}{\kw{VertexRDD}}

\newcommand{\devevaltitle}{OSS-UAgent\space}
\newcommand{\deveval}{\kw{OSS}-\kw{UAgent}}
\newcommand{\researcher}{\kw{Researcher}}
\newcommand{\developer}{\kw{Developer}}
\newcommand{\evaluator}{\kw{Evaluator}}
\newcommand{\coder}{\kw{Code~Generator}}

\newcommand{\ldbcdg}{\kwnospace{LDBC}-\kw{DG}}
\newcommand{\stgt}{\kw{S3G2}}
\newcommand{\ldbc}{\kw{LDBC~Graphalytics}}
\newcommand{\graphfive}{\kw{Graph500}}
\newcommand{\fftdg}{\kw{FFT}-\kw{DG}}
\title{\devevaltitle: An Agent-based Usability Evaluation Framework for Open Source Software}

\author{Lingkai Meng}
\affiliation{
  \institution{Shanghai Jiao Tong University}
}
\email{mlk123@sjtu.edu.cn}

\author{Yu Shao}
\affiliation{
  \institution{East China Normal University}
}
\email{yushao@stu.ecnu.edu.cn}

\author{Long Yuan}
\authornote{Corresponding author.}
\affiliation{
  \institution{Wuhan University of Technology}
}
\email{longyuanwhut@gmail.com}

\author{Longbin Lai}
\affiliation{
  \institution{Alibaba Group}
}
\email{longbin.lailb@alibaba-inc.com}

\author{Peng Cheng}
\affiliation{
  \institution{Tongji University}
}
\email{cspcheng@tongji.edu.cn}

\author{Wenyuan Yu}
\affiliation{
  \institution{Alibaba Group}
}
\email{wenyuan.ywy@alibaba-inc.com}

\author{Wenjie Zhang}
\affiliation{
  \institution{University of New South Wales}
}
\email{wenjie.zhang@unsw.edu.au}

\author{Xuemin Lin}
\affiliation{
  \institution{Shanghai Jiao Tong University}
}
\email{xuemin.lin@gmail.com}

\author{Jingren Zhou}
\affiliation{
  \institution{Alibaba Group}
}
\email{jingren.zhou@alibaba-inc.com}

\begin{abstract}

Usability evaluation is critical to the impact and adoption of open source software (OSS), yet traditional methods relying on human evaluators suffer from high costs and limited scalability. To address these limitations, we introduce \deveval, an automated, configurable, and interactive agent-based usability evaluation framework specifically designed for open source software. Our framework employs intelligent agents powered by large language models (LLMs) to simulate developers performing programming tasks across various experience levels (from Junior to Expert). By dynamically constructing platform-specific knowledge bases, \deveval ensures accurate and context-aware code generation. The generated code is automatically evaluated across multiple dimensions, including compliance, correctness, and readability, providing a comprehensive measure of the software's usability. Additionally, our demonstration showcases \deveval's practical application in evaluating graph analytics platforms, highlighting its effectiveness in automating usability evaluation.
\end{abstract}

\maketitle

\pagestyle{\vldbpagestyle}
\begingroup\small\noindent\raggedright\textbf{PVLDB Reference Format:}\\
\vldbauthors. \vldbtitle. PVLDB, \vldbvolume(\vldbissue): \vldbpages, \vldbyear.\\
\href{https://doi.org/\vldbdoi}{doi:\vldbdoi}
\endgroup
\begingroup
\renewcommand\thefootnote{}\footnote{\noindent
This work is licensed under the Creative Commons BY-NC-ND 4.0 International License. Visit \url{https://creativecommons.org/licenses/by-nc-nd/4.0/} to view a copy of this license. For any use beyond those covered by this license, obtain permission by emailing \href{mailto:info@vldb.org}{info@vldb.org}. Copyright is held by the owner/author(s). Publication rights licensed to the VLDB Endowment. \\
\raggedright Proceedings of the VLDB Endowment, Vol. \vldbvolume, No. \vldbissue\ %
ISSN 2150-8097. \\
\href{https://doi.org/\vldbdoi}{doi:\vldbdoi} \\
}\addtocounter{footnote}{-1}\endgroup

\ifdefempty{\vldbavailabilityurl}{}{
\vspace{.3cm}
\begingroup\small\noindent\raggedright\textbf{PVLDB Artifact Availability:}\\
The source code, data, and/or other artifacts have been made available at \url{https://github.com/Lingkai981/OSS-UAgent}.
\endgroup
}

\section{Introduction}

The usability of open source software (OSS) is a qualitative characteristic evaluating its ease of learning and usage, significantly impacting their adoption, development efficiency, and user satisfaction~\cite{andreasen2006usability,DBLP:conf/icse/Myers17,DBLP:journals/cacm/MyersS16,DBLP:journals/csr/RaufTP19}.
A highly usable OSS platform helps efficient task completion, enhances developer productivity, reduces error likelihood, promotes broader adoption within developer communities, improves code maintainability, and fosters a satisfying development experience. 
However, despite its importance, conducting usability evaluations remains challenging in practice. Traditional evaluation methods primarily rely on empirical studies involving human participants, where developers are recruited to interact with the platform and answer interview questions~\cite{farooq2010api,piccioni2013empirical}.
However, these human-centric methods face substantial challenges, including high costs, limited scalability, and difficulties in recruiting qualified participants, particularly when evaluating OSS with extensive functionality and complex application scenarios~\cite{farooq2010api}.

\begin{figure*}[t]
	\begin{center}
		\begin{tabular}[t]{c}
			\includegraphics[width=2\columnwidth]{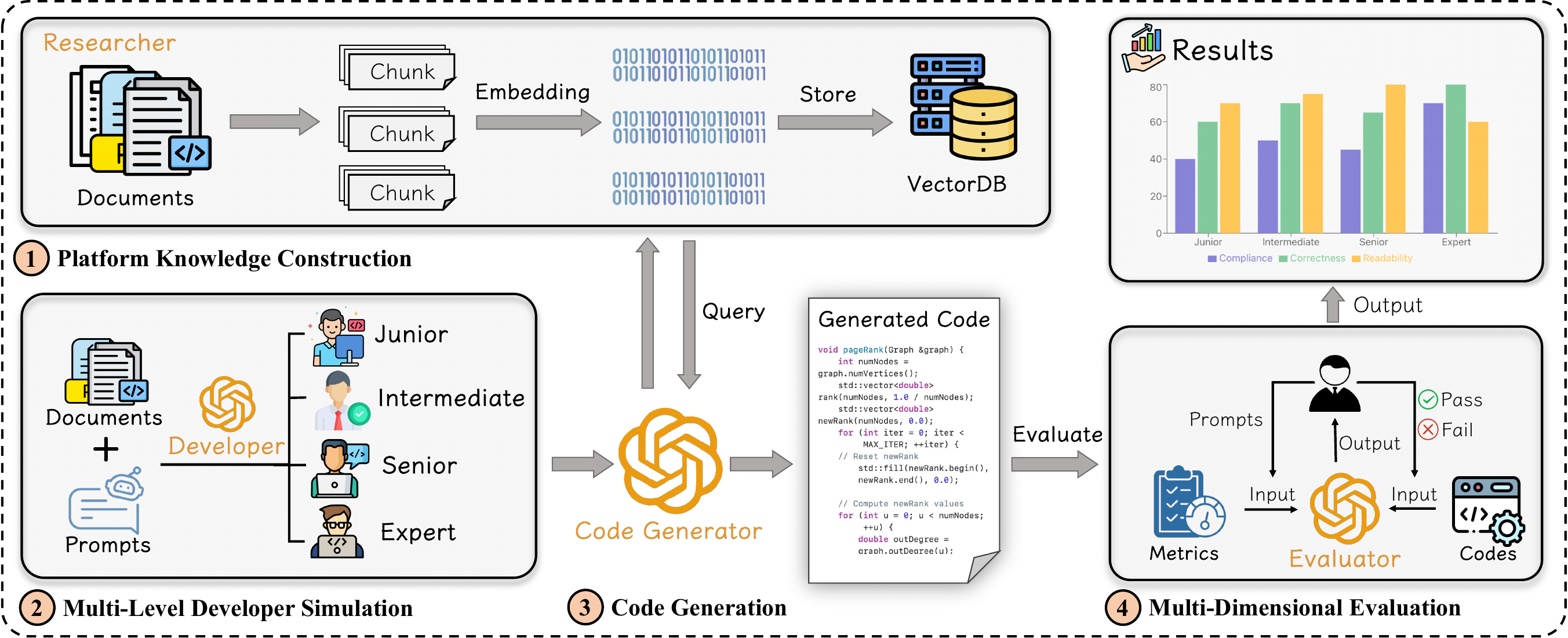}
		\end{tabular}
	\end{center}
	 \vspace{-1em}
	\caption{Agent-based Usability Evaluation Framework}
	\label{fig:llm_framework}
	 \vspace{-0.8em}
\end{figure*}

To address the shortcomings of existing evaluation methods, we introduce \deveval, a scalable and automated usability evaluation framework that replaces human evaluators with intelligent agents powered by large language models (LLMs), which have demonstrated capabilities comparable to, or even exceeding, human performance in various programming and evaluation tasks, making them a feasible solution to reduce evaluation costs and improve scalability~\cite{jiang2024learners,chiang2023can,wang2024understanding}.
Specifically, our framework first collects platform-specific data, including API documentation, research papers, and sample code, to construct a vectorized knowledge base. This knowledge base enables context-aware retrieval of relevant information during code generation, ensuring alignment with platform standards.
\deveval simulates developers across multiple experience levels (from Junior to Expert) by generating role-specific prompts that reflect varying platform familiarity.
Subsequently, the LLM-based agent generates corresponding code based on these multi-level prompts with knowledge base support. Finally, the generated code is evaluated against standard implementations based on compliance, correctness, and readability metrics.
\deveval significantly reduces evaluation costs and enhances scalability by automating the evaluation process, enabling the evaluation of large-scale OSS platforms without the need for extensive human participation.


As graph analytics platforms have been widely studied in the database community, we use them as representative examples to demonstrate our \deveval through an interactive graphical user interface\footnote{The code can be accessed at \url{https://github.com/Lingkai981/OSS-UAgent}.}. This includes automatic retrieval and analysis of data from platform GitHub repositories, construction of a dynamic knowledge base, generation of multi-level prompts, and automated code evaluation. To ensure fairness and consistency in assessments, the collected platform data will be anonymized. The user interface presents generated code, evaluation outcomes, and detailed reports segmented by experience level, providing clear insights into the platform’s usability.

\section{Framework Overview}

The core idea of our usability evaluation framework \deveval is to leverage LLM-based agents to simulate developers with different experience levels, enabling them to complete platform-specific tasks and to evaluate the usability of the OSS by analyzing the quality of the generated code.
As shown in Figure~\ref{fig:llm_framework}, \deveval consists of multiple agents and has four main steps, \textbf{Platform Knowledge Construction}, \textbf{Multi-Level Developer Simulation}, \textbf{Code Generation} and \textbf{Multi-Dimensional Evaluation}. Specifically, $(1)$ the \underline{\researcher} agent automatically gathers and processes platform-specific documents, then builds a vector database that supports efficient, context-aware retrieval for code generation; $(2)$ the \underline{\developer} agent simulates developers at different experience levels (Junior, Intermediate, Senior, Expert); $(3)$ the \underline{\coder} agent generates code based on prompts and retrieved knowledge; and $(4)$ the \underline{\evaluator} agent scores the generated code according to predefined criteria.


\subsection{Platform Knowledge Construction}

To ensure the generated code follows platform-specific standards, our framework introduces the \researcher agent, which dynamically builds and maintains a structured knowledge base.
As illustrated in Figure~\ref{fig:llm_framework}~\circled{1}, \researcher first gathers platform-specific documents, including research papers, API documentation, sample code, and coding guidelines. 
These documents are preprocessed into smaller chunks, embedded into vector representations, and stored in a vector database to support efficient similarity-based retrieval.

During code generation, the \coder queries the vector database to retrieve the most relevant data, which is injected into the prompt context. This ensures the generated code follows platform-specific standards and improves overall accuracy.

To ensure fairness and eliminate bias, we anonymize all platform-specific identifiers (e.g., platform names, unique function names, and parameter identifiers) during the preprocessing phase. This anonymization ensures that \evaluator evaluates platforms based on general usability rather than prior familiarity with specific platform details.



\subsection{Multi-Level Developer Simulation}


A platform’s usability can vary significantly depending on a developer’s experience. However, it is not straightforward to determine which experience level a generic LLM might represent in practice. To address this challenge, our framework employs the \developer agent to simulate developers at four distinct experience levels, Junior, Intermediate, Senior, and Expert, through hierarchical prompts, ensuring a more comprehensive and realistic evaluation.


\begin{itemize} [leftmargin=*]
	\item \textbf{Level 1 (Junior)} represents individuals with basic programming knowledge but little experience with APIs. Prompts provide only task descriptions.
	
	\item \textbf{Level 2 (Intermediate)} corresponds to developers with some exposure to APIs but limited expertise. Prompts include general guidance, such as function names and parameter hints.
	
	\item \textbf{Level 3 (Senior)} represents experienced developers familiar with best practices and efficient API usage. Prompts provide more structured information, including example use cases.
	
	\item \textbf{Level 4 (Expert)} simulates highly skilled developers who can understand complex requirements and optimize solutions. Prompts contain comprehensive details and and they expect high-quality, efficient implementations.
\end{itemize}

For each experience level, \developer receives tailored prompts, such as:
Junior Prompt: ``\textit{You are a beginner developer and you have no prior knowledge of the API in platform X ... Here is the relevant information about the task...}''
Expert Prompt: ``\textit{You are an expert developer ... Below is detailed information about the API and related concepts...}''

\subsection{Code Generation}
\label{sec:coder}


In this step (Figure~\ref{fig:llm_framework}~\circled{3}), the \coder agent receives multi-level prompts from \developer. Each prompt is linked to a specific developer role and represents a distinct experience level. Based on the task requirements and the developer’s role, \coder queries the vector database built by \researcher for relevant API function details. It then adds this information to the prompt to ensure the generated code follows platform standards. Finally, \coder produces a set of code implementations that match the style and knowledge of each role. These implementations become the main input for the evaluation described in Section~\ref{sec:Evaluation}.

\subsection{Multi-Dimensional Evaluation}
\label{sec:Evaluation}
Determining the quality of the generated code is critical for evaluating the usability of developer platforms. Correctness and readability are widely recognized metrics in API usability evaluation~\cite{DBLP:conf/icse/Myers17,piccioni2013empirical,DBLP:journals/csr/RaufTP19}. Correctness reflects whether the API documentation and examples sufficiently guide developers to achieve intended functionality accurately. Readability measures whether the API's design encourages clear and maintainable code.
However, we observe that LLMs often exhibit ``hallucination'', focusing on general programming patterns or inventing nonexistent API functions while ignoring platform-specific APIs, typically due to unclear or ambiguous prompts. This limitation mirrors the behavior of human developers: less experienced programmers are more likely to make mistakes when dealing with poorly designed APIs. 
To address this, we introduce a new metric, compliance, which measures how closely the generated code aligns with standard code. This metric reflects the API's intuitiveness and accessibility for developers with varying skill levels. By assessing whether the API enables users to easily produce code that establishes best practices and standards, compliance provides an objective basis for scoring API usability.


The details of evaluation metrics are as follows:

\begin{itemize}[leftmargin=*]
	\item \textbf{Compliance.}  Checking the adherence to platform-specific coding standards and best practices by comparing the generated code with standard code examples. Compliance ensures that the code integrates well with existing systems
	
	\item \textbf{Correctness.} This metric is for ensuring the generated code performs the intended task accurately. This includes verifying the logic of the code and the correctness of function calls.
	
	\item \textbf{Readability.} This metric focuses on code clarity and maintainability. Readable code is easier to understand, modify, and debug. It should be well-structured, logically grouped, and follow consistent naming conventions.
\end{itemize}

Figure~\ref{fig:llm_framework}\circled{4} illustrates the training process: we first provide detailed scoring criteria and set basic requirements instructions to get \evaluator. Then, we introduce some test code and provide feedback based on the output results to optimize the \evaluator's instructions. We iterate this process until it can produce stable and satisfactory evaluation results.




\section{Demonstration for Graph Analytics Platforms}
\label{sec3}

\begin{figure*}[h]
	\begin{center}
		\begin{tabular}[t]{c}
			\includegraphics[width=2\columnwidth]{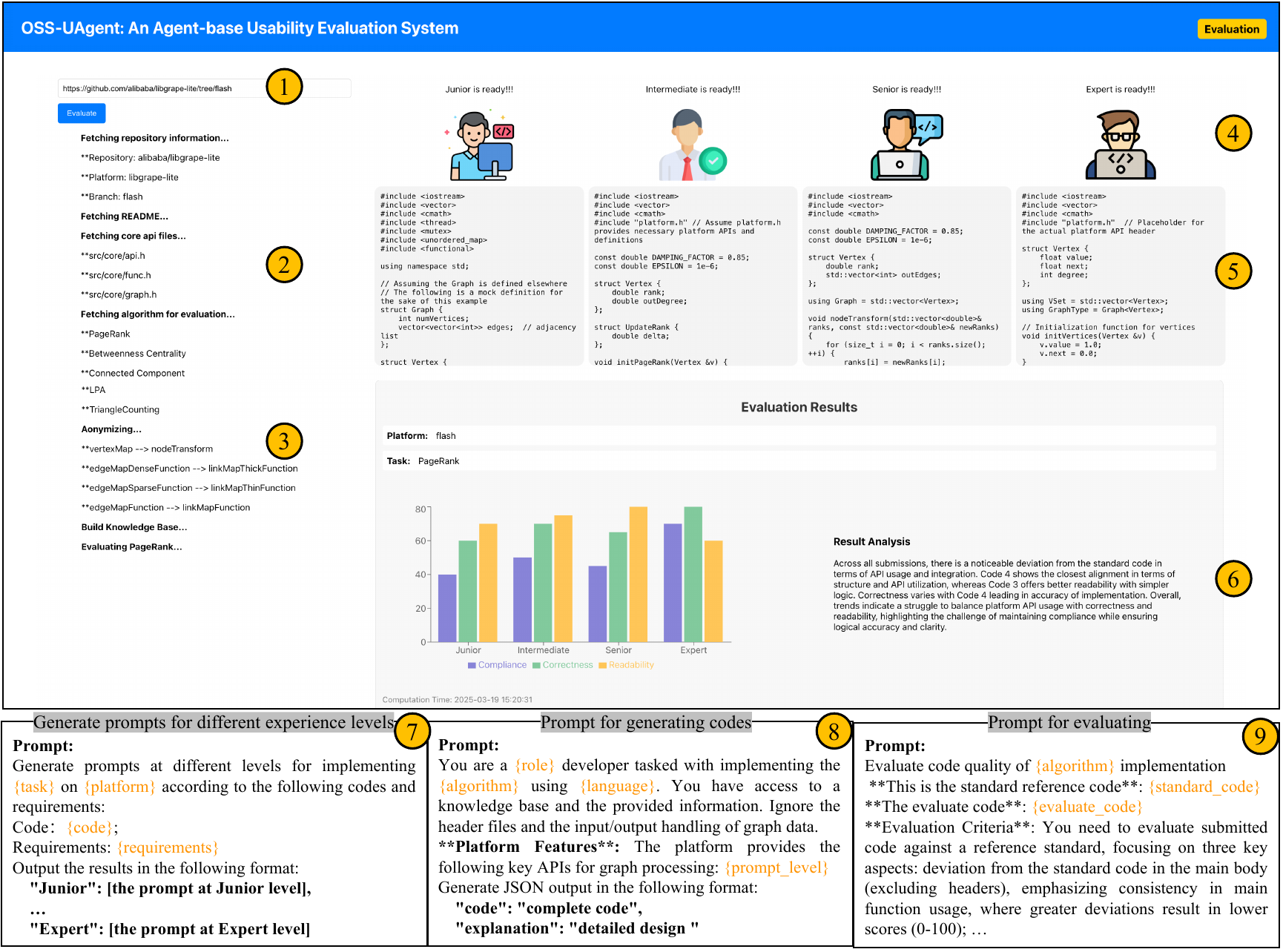}
		\end{tabular}
	\end{center}
	 \vspace{-1em}
	\caption{An Example Workflow of Using \devevaltitle’s GUI}
	\label{fig:Demo2}
	 \vspace{-0.8em}
\end{figure*}


In this section, we demonstrate our usability evaluation system using graph analytics platforms as an example.
Graph analytics platforms have received significant attention from the database community due to their efficiency in data management and analysis, making them ideal candidates for demonstrating our \deveval.
The system is built upon our agent-based usability evaluation framework and offers an intuitive graphical user interface. The backend is developed using Python, and the frontend is constructed with JavaScript and the React library.
Users only need to provide the GitHub repository URL of the target platform. From there, the system automates the entire evaluation process, including data retrieval, knowledge base construction, code generation, and comprehensive multi-dimensional evaluation.

Although our demonstration focuses on graph analytics platforms, the underlying framework is highly generalizable and can be effectively applied to other open source software, such as databases, machine learning frameworks, and development platforms.

\stitle{Platform Data Retrieval \& Knowledge Base Construction.}
The user first inputs the target evaluation platform’s GitHub URL (Figure~\ref{fig:Demo2}~\circled{1}). The system then automatically fetches basic information, including the README, core API files, and example code. Due to varying naming conventions across platforms, locating core API and example code files directly is challenging. To address this, the LLM-based \researcher agent parses all file paths and filenames to automatically identify the core API and example code files, as shown in Figure~\ref{fig:Demo2}~\circled{2}.

To ensure evaluation fairness and anonymity, the platform and core function names are anonymized. This anonymization process is also conducted by the \researcher, which generates a set of uniform anonymization rules based on the retrieved data (Figure~\ref{fig:Demo2}~\circled{3}). All subsequent processes, including knowledge base construction and code generation, apply these rules consistently. Finally, the anonymized README content and API documentation are segmented, vectorized, and stored in a vector database, forming a structured knowledge base.

\stitle{Code Generation.} Based on the fetched files, our system automatically generates tailored prompts corresponding to four developer experience levels (Junior, Intermediate, Senior, Expert). These prompts guide the \coder to generate code that adheres to platform-specific APIs and best practices (Figure~\ref{fig:Demo2}\circled{4}). The detailed template for these prompts is illustrated in Figure\ref{fig:Demo2}~\circled{7}. The requirements for each experience level are defined as follows:

\begin{itemize} [leftmargin=*]
	\item \textbf{Level 1 (Junior).} At this level, no specific technical details are provided. The prompt only consists of a description of the task, without any guidance on how to implement it.
	
	\item \textbf{Level 2 (Intermediate).} This level offers minimal technical information to guide the code generation. Basic prompts are given, including the names of core APIs and parameters.
	
	\item \textbf{Level 3 (Senior).} This level provides detailed usage instructions for the relevant APIs, including the names of the APIs and parameters, and a detailed introduction to them. In addition, some example code is provided to guide the usage of API functions. 
	
	\item \textbf{Level 4 (Expert).} In addition to the detailed API instructions similar to the previous level, this level also includes the pseudo-code of the relevant algorithm.
\end{itemize}

Once the prompts are generated, the system provides them to the \coder, which then produces code implementations corresponding to each experience level. The generated code is displayed in the evaluation interface, allowing direct comparison across different experience levels (Figure~\ref{fig:Demo2}~\circled{5}). The prompt template for generating code is shown in Figure~\ref{fig:Demo2}~\circled{8}.

\stitle{Code Evaluation \& Result Presentation.}
The generated code is assessed against a standard reference implementation based on three key criteria, Compliance, Correctness, and Readability, detailed in Section~\ref{sec:Evaluation}.
Each generated code is scored based on these criteria, and the results are presented in a visual format for direct comparison across different experience levels (Figure~\ref{fig:Demo2}~\circled{6}). The evaluation results provide insights into the quality of generated code at different experience levels, helping users to clearly understand API usability and developer-friendliness.

To ensure fair and consistent assessment, the system employs a structured prompt to guide the \evaluator during evaluation. The prompt template is illustrated in Figure~\ref{fig:Demo2}~\circled{9}. Based on this prompt, the system generates a detailed evaluation report that highlights key differences across generated code.

\bibliographystyle{ACM-Reference-Format}
\bibliography{main}

\end{document}